\documentclass{ws-p8-50x6-00_mod}

\begin{document}

\title{Determination of the strong coupling constant using subjet multiplicities in 
Neutral Current Deep Inelastic Scattering}

\author{O. Gonz\'alez}

\address{{\bf (on behalf of the ZEUS Collaboration)}\\
Dpto. de F\'{\i}sica Te\'orica, Universidad Aut\'onoma de Madrid, Cantoblanco, \\
E-28049 Madrid, Spain \\
E-mail: gonzalez@mail.desy.de}


\maketitle

\abstracts{The internal structure of the jets produced 
in neutral current interactions 
for $Q^2>125$~GeV$^2$ has been studied  
using the subjet multiplicity with the ZEUS detector at HERA. 
Jets are identified in the laboratory frame 
by applying the longitudinally invariant $k_T$-cluster algorithm.
Next-to-leading order QCD calculations have been obtained and compared to the data;
a good agreement has been observed in the region where hadron-to-parton corrections are
small, $E_{T,jet}> 25$~GeV. 
In this region, the strong coupling constant is extracted
and the resulting value is $\alpha_s (M_Z)=0.1185 \pm 0.0016 \; \rm{(stat.)} 
^{+0.0067}_{-0.0048} \; \rm{(syst.)} ^{+0.0089}_{-0.0071} \; \rm{(th.)}$.}

\section{Introduction}

The study of the internal structure of jets gives insight into the transition between partons
produced in a hard process and the experimentally observable spray of hadrons. 
This type of analysis is usually made using jet shapes\cite{ref_jetshp}, where the energy
flow inside the jet is considered, or the subjet multiplicity\cite{ref_sbj}, where the study is
done with jet-like structures~(subjets) within a given jet. These structures are called subjets and 
are formally defined as the original jets, but resolved at a smaller scale.
At sufficiently high jet energy, where fragmentation effects are small, the internal structure of 
the jets is expected to depend mainly on the parton-radiation pattern and should be calculable 
in pertubative QCD.

Studies of the internal structure of jets have been made in $p\bar{p}$ collisions at 
Tevatron\cite{ref_tevasbj} and in $e^+e^-$ interactions at LEP\cite{ref_lepsbj}. It has been 
observed that phenomenological models for the parton radiation based on QCD give
a good description of the data. In addition, the results are in agreement with the 
expectation that the internal structure is determined by the type of the primary parton. It is 
observed that gluon-initiated jets are broader (i.e. contain a larger number of subjets) than
quark-initiated jets, as predicted by QCD from the larger color charge of the gluon. At HERA,
measurements of subjet multiplicities have previously been presented in quasi-real
photon proton collisions (photoproduction)\cite{ref_phpsbj} and in neutral current deep 
inelastic scattering\cite{ref_h1sbj}. In photoproduction, the mean subjet multiplicity 
($<n_{sbj}>$) is observed to become larger as the jet pseudorapidity ($\eta_{jet}$) 
increases, in agreement with the predicted increase in the fraction of gluon-initiated jets.
In deep inelastic scattering for jets defined in the Breit frame, 
it has been observed that jets 
pcontain a smaller number of subjets with increasing transverse energies and towards the forward
region of the Breit frame.

New measurements of the mean subjet multiplicity 
for jets produced in neutral current deep 
inelastic scattering with the ZEUS detector at HERA are presented here.  
The jets are selected using the longitudinally invariant 
$k_T$-cluster algorithm\cite{ref_kt} in the laboratory frame. 
Next-to-leading order (NLO) QCD calculations 
for $<n_{sbj}>$ are now available for jets defined in this frame and they are compared
to the data after hadronization corrections are applied.
From this comparison, the value of the strong 
coupling constant is extracted.

\section{Measurement of the mean subjet multiplicity}

The data sample used in this analysis was collected with the ZEUS detector at HERA and corresponds to 
an integrated luminosity of 38~pb$^{-1}$.
During 1996-1997, HERA operated with positrons of energy $E_e = 27.5$~GeV 
colliding with protons of energy $E_p=820$~GeV. 
Neutral current events were selected by requiring a positron 
candidate identified in the main calorimeter (UCAL).  After removing the background contributions, events 
with $Q^2>125$~GeV$^2$ were kept and the jet algorithm was applied to the UCAL cells
belonging to the hadronic system. The jet search is performed in the pseudorapidity 
($\eta$)~\footnote{The pseudorapidity is defined as $\eta=-\ln (\tan (\theta / 2))$ where $\theta$ is the 
polar angle with respect to the proton beam direction.} - azimuth ($\phi$) plane of the laboratory frame. The
jet variables are defined according to the Snowmass convention\cite{ref_snowmass}.
The inclusive sample of jets with transverse energy $E_{T,jet} > 15$~GeV and $-1 < \eta_{jet} < 2$
has been studied.

For each jet in the sample, the jet algorithm was re-applied to
all the particles assigned to the given jet. The clustering 
was stopped when the distances $y_{ij}$ between all 
pair of particles $i, j$ are above some 
resolution scale 
$y_{cut}$
\begin{displaymath}
          y_{ij} = \frac{ min\{ E_{T,i},E_{T,j}\}^2}{E_{T,jet}^2} \; 
                      \left( \Delta \eta_{ij}^2 +   \Delta \phi_{ij}^2 \right) \; > \; y_{cut}.
\end{displaymath}
All clusters found in this way are called subjets.

The quantity  $\sqrt{y_{cut}} E_{T,jet} $ 
determines the scale at which the internal structure of the jet is resolved. If this quantity is large, 
then the subjet structure is given by parton-radiation processes in which the relative
transverse momentum is hard enough and perturbative QCD should be applicable to describe
the subjet structure of the jets.

The mean subjet multiplicity is defined 
as the average of the number of subjets in the considered sample of jets at a given $y_{cut}$. The
mean subjet multiplicity has been studied as a function of $y_{cut}$ and of the jet variables for
a fixed $y_{cut}=0.01$.

A simulation of the detailed properties of the hadronic final state is available 
in various Monte Carlo
event generators. The generated samples of events have been passed through a 
detector simulation and 
analysed by the same program chain as the data. 
The response of the detector to jets of hadrons and
the correction factors for $<n_{sbj}>$ have been determined by 
applying the same jet algorithm
to the hadron level of the simulated events.

\begin{figure}[t]
  \hspace{-0.5cm}
    \epsfxsize=8cm 
    \epsfbox{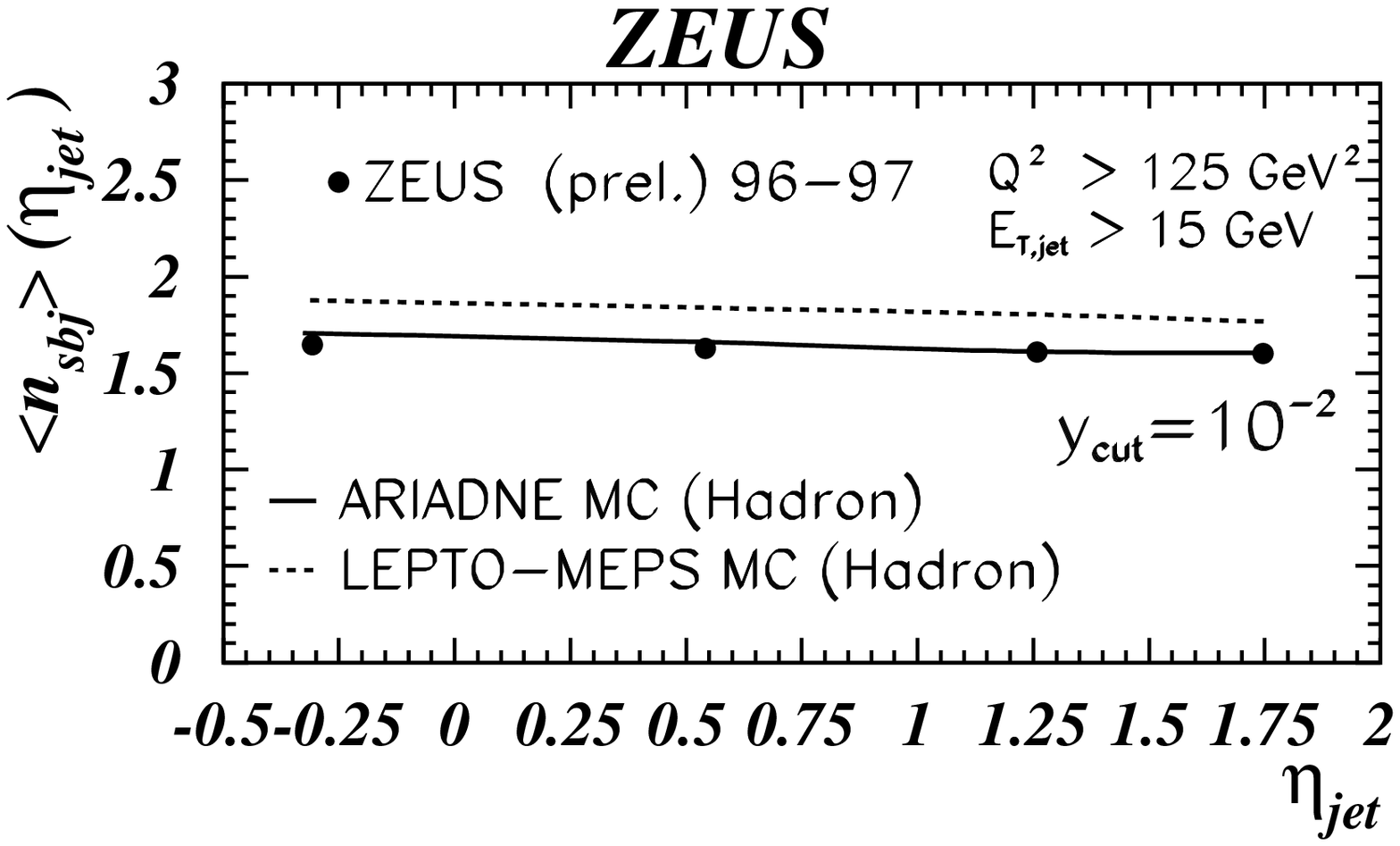} 
    \epsfxsize=7cm 
    \epsfbox{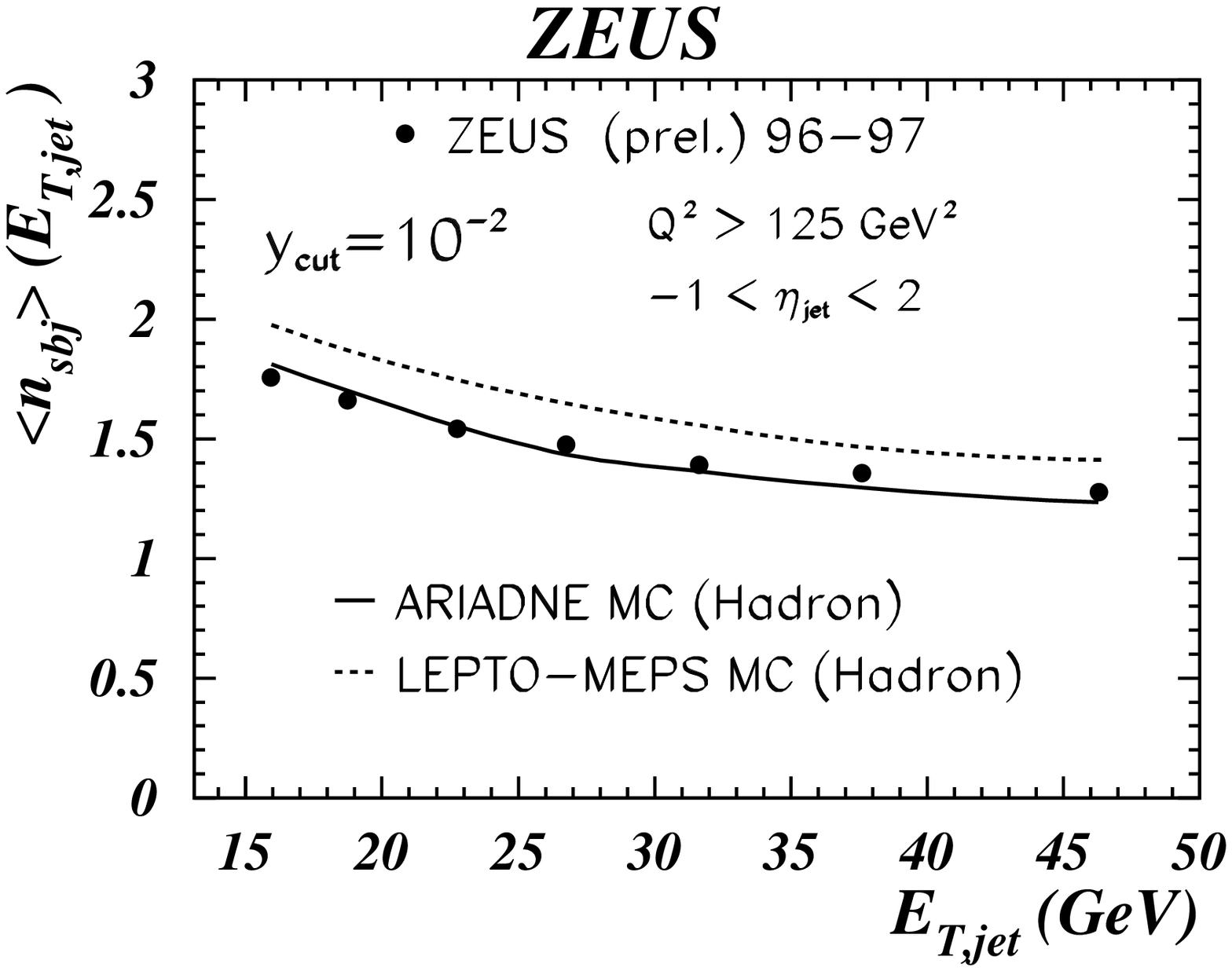} 
 \vspace{-1.cm}
    \caption{Measured mean subjet multiplicity corrected to hadron level for jets produced
                  in neutral current DIS interactions as a function (left) 
                  of the 
                  jet pseudorapidity and (right) 
                  of the jet transverse energy in the laboratory
                  frame. \label{fig:hadron}}
\end{figure}
                
Figure~\ref{fig:hadron} shows the measurements of the mean subjet multiplicity as a 
function of the jet 
pseudorapidity and jet transverse energy for $y_{cut}=0.01$. 
The data have been corrected to the hadron level and compared to the predictions of two Monte 
Carlo models.
The color-dipole model as implemented in ARIADNE\cite{ref_ari} 
gives a very good description of the data
while the matrix element plus parton shower (MEPS) model of LEPTO\cite{ref_lepto} is 
above the data. 

It is observed that the mean subjet multiplicity has no $\eta_{jet}$ dependence. 
This is
different than the results observed in photoproduction\cite{ref_phpsbj},  
where an increase is observed as $\eta_{jet}$ increases. 
These two different behaviors can be understood 
in terms of the admixtures of quark- and gluon-initiated jets
in the two reactions. In the analysis of neutral current deep inelastic
scattering presented here, the observed distribution
is consistent with the expectation that the sample is dominated by quark-initiated
jets from the Born process for the entire $\eta_{jet}$ range. Comparisons with the predictions
by the Monte Carlo models for quark- and gluon-initiated jets confirm this explanation.

On the other hand, the mean subjet multiplicity is observed to decrease as the transverse energy 
of the jets increases. 
This means that the jets become narrower as $E_{T,jet}$ increases, in 
agreement with the expectations. A similar behavior has been observed in
previous results at HERA, Tevatron and LEP by using subjet multiplicities or jet shapes.

\section{NLO QCD calculations}

For jets defined in the laboratory frame, it 
is possible to obtain predictions for
the mean subjet multiplicity from perturbative QCD to the second order in $\alpha_s$. 
The perturbative QCD calculations have been obtained with the program DISENT\cite{ref_disent}.
The number of flavors was set to 5 and the renormalization~($\mu_R$) and factorization~($\mu_F$)
scales were both set to $Q$. The main source of theoretical uncertainty is due to terms beyond
NLO and has been estimated by varying $\mu_R$ between $Q/2$ and $2\,Q$, keeping $\mu_F$ fixed
at $Q$.

\begin{figure}[t]
    \hspace{-1.cm}
    \epsfxsize=7.5cm 
    \epsfbox{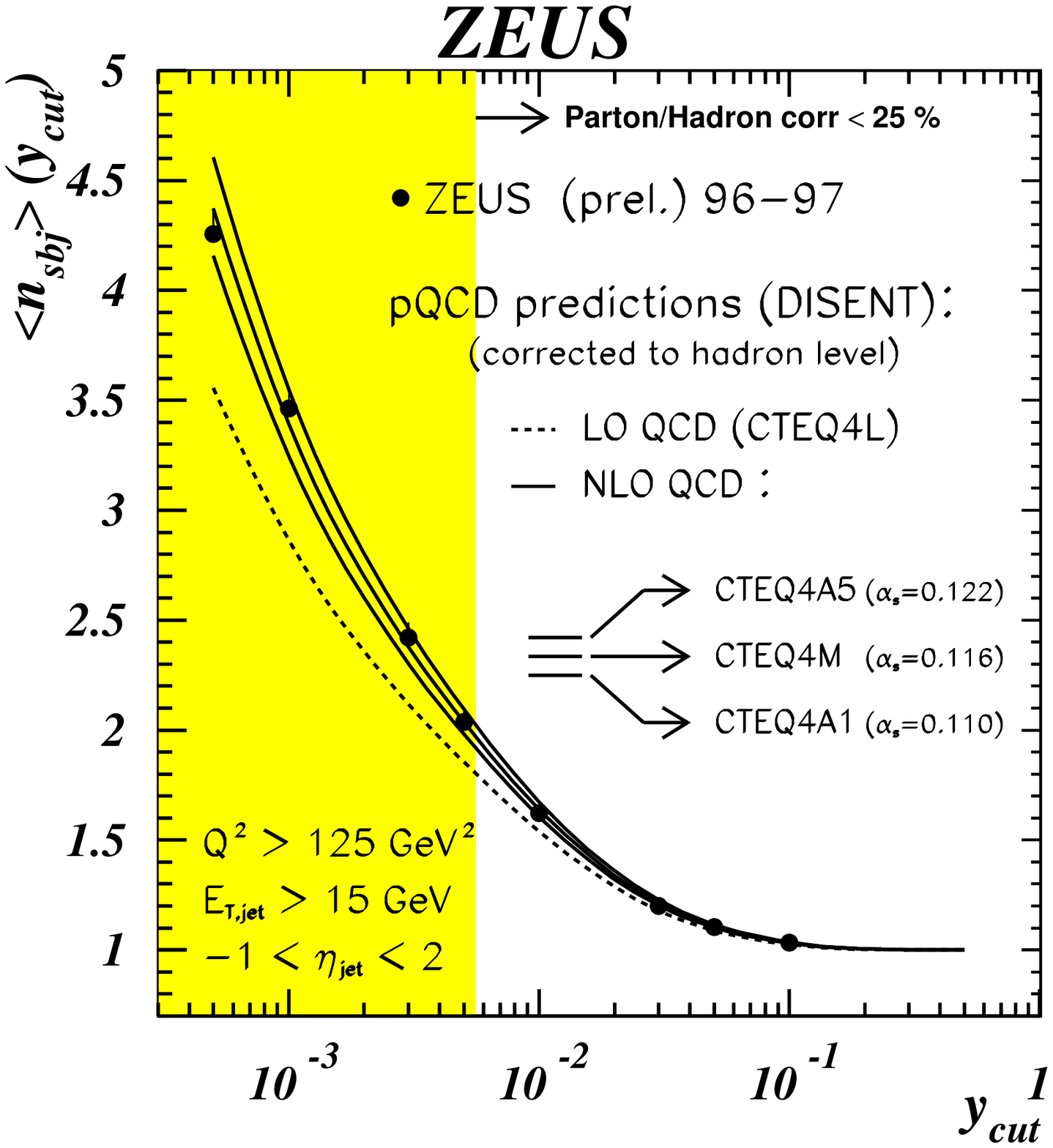} 
    \hspace{0.5cm}
    \epsfxsize=7.5cm 
    \epsfbox{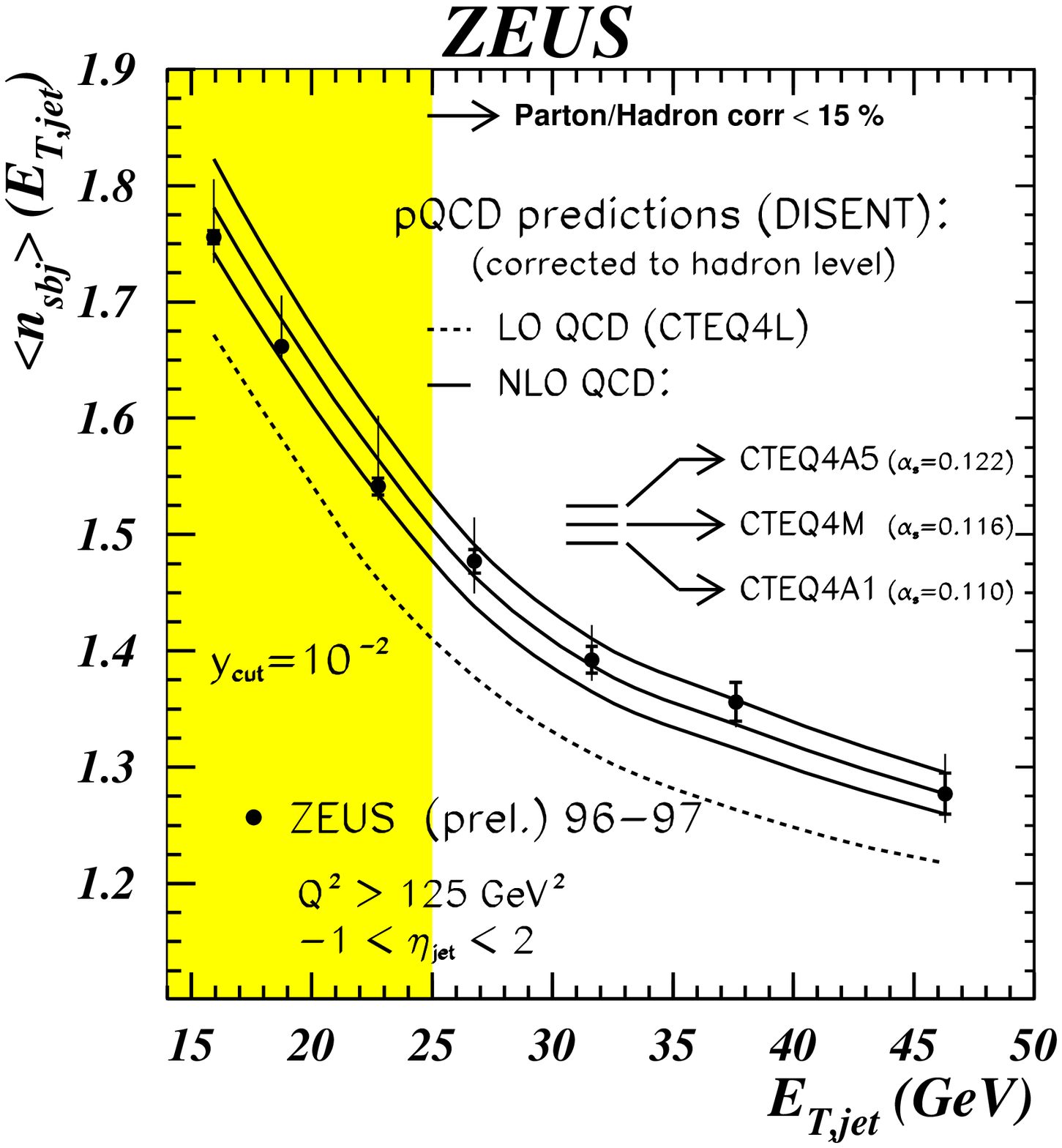}
  \vspace{-.5cm}
    \caption{Measured mean subjet multiplicity corrected to hadron level 
             for jets produced
                  in neutral current DIS interactions.  The data are 
             compared to NLO QCD calculations
                  for different values of $\alpha_s$:  (left plot) 
                  $<n_{sbj}>$ as a 
                  function of $y_{cut}$; (right plot) $<n_{sbj}>$ at $y_{cut}=0.01$ as a 
                  function of the transverse energy of the jets 
                  in the laboratory frame. \label{fig:nlo}}
\end{figure}

To compare with the data, the predictions given by DISENT (at parton level)
have been corrected for hadronization effects
by using the Monte Carlo models. In the ARIADNE and LEPTO models, the hadronization 
is performed by using the LUND string model\cite{ref_lund} as implemented
in JETSET\cite{ref_jetset}. 

The perturbative QCD predictions corrected for hadronization effects 
are compared to the data in figure~\ref{fig:nlo}. 
The predictions give a good description of the data, even where parton-to-hadron corrections
are large. NLO QCD predictions have been obtained for different values of 
$\alpha_s$. 
The parametrizations of the proton parton distribution 
functions assuming different values of $\alpha_s (M_Z)$ as given in the 
CTEQ4A-series\cite{ref_cteq4}  
have been used; this procedure takes into account the correlations
between the value of $\alpha_s (M_Z)$ used 
in the matrix elements and that assumed in the parametrization of the
parton distribution functions. 

The comparison of the data with the theoretical predictions for different values of $\alpha_s$
shows that the measurements are sensitive to $\alpha_s(M_Z)$.

\section{Determination of $\alpha_s$}

The value of $\alpha_s$ 
has been extracted in the region where parton-to-hadron corrections are below 15\%.
The method consists of
making a parametrization of the dependence on $\alpha_s$ of the theoretical
predictions for the mean subjet multiplicity. 
A preliminary study of the systematic uncertainties has been performed; the most important
contribution is the uncertainty on the modelling of the hadronic final state.

The extracted value of $\alpha_s (M_Z)$ using $< n_{sbj} >$ at $y_{cut}=0.01$ and for jets
with $E_{T,jet} > 25$~GeV is

\vspace{.3cm}

\centerline{$\alpha_s (M_Z)=0.1185 \pm 0.0016 \; \rm{(stat.)} 
^{+0.0067}_{-0.0048} \; \rm{(syst.)} ^{+0.0089}_{-0.0071} \; \rm{(th.)}$}

\vspace{.5cm}

\noindent
in good agreement with the world average value\cite{ref_alphas}.


\begin{thebibliography}{99}

\bibitem{ref_jetshp} S.D. Ellis, Z. Kunszt and D.E. Soper \Journal{\PRL}{69}{3615}{1992}.

\bibitem{ref_sbj} S. Catani et al., \Journal{\NPB}{377}{445}{1992} and \Journal{\NPB}{383}{419}{1992}; 
     M.H. Seymour, \Journal{\NPB}{421}{545}{1994} and  \Journal{\PLB}{378}{279}{1996}. 

\bibitem{ref_tevasbj} CDF Collab., F. Abe et al., \Journal{\PRL}{70}{713}{1993}.\\
D0 Collab., S. Abachi et al., \Journal{\PLB}{357}{500}{1995}.\\
D0 Collab., V.M. Abazov et al., preprint FERMILAB-PUB-01-248-E (2001),
hep-ex/0108054.

\bibitem{ref_lepsbj} ALEPH Collab., D. Buskulic et al., \Journal{\PLB}{346}{389}{1995}.

\bibitem{ref_phpsbj} ZEUS Collab. {\it Contributed paper to ICHEP'99 (Tampere, July 1999) N-530}.

\bibitem{ref_h1sbj} H1 Collab., C. Adloff et al., \Journal{\NPB}{545}{3}{1999}.

\bibitem{ref_kt} S. Catani et al., \Journal{\NPB}{406}{187}{1993}; 
                          S.D. Ellis and D.E. Soper, \Journal{\PRD}{48}{3160}{1993}.

\bibitem{ref_snowmass} J. Huth et al. Proc. of the 1990 DPF Summer Study on High Energy Physics, 
                                      Snowmass, Colorado, edited by E.L. Berger (World Scientific, Singapore, 1992) p. 134.

\bibitem{ref_ari} L. L\"onnblad, \Journal{{\em Comp. Phys. Comm.}}{71}{15}{1992}; version used: 4.08.

\bibitem{ref_lepto} G. Ingelman, A. Edin and J. Rathsman, 
                              \Journal{{\em Comp. Phys. Comm.}}{101}{108}{1997}; version used: 6.5.

\bibitem{ref_disent} S. Catani and M.H. Seymour, \Journal{\NPB}{485}{291}{1997}.

\bibitem{ref_lund} B. Andersson et al., \Journal{{\em Phys. Rep.}}{97}{31}{1983}.

\bibitem{ref_jetset} T. Sj\"ostrand, \Journal{{\em Comp. Phys. Comm.}}{39}{347}{1986};
                               T. Sj\"ostrand and M. Bengtsson, \Journal{ibid.}{43}{367}{1987}.

\bibitem{ref_cteq4} H.L. Lai et al., \Journal{\PRD}{55}{1280}{1997}.

\bibitem{ref_alphas} Particle Data Group, D.E. Groom et al., \Journal{{\em Eur. Phys. J.}}{C 15}{1}{2000};
                                 S. Bethke, \Journal{{\em J. Phys.}}{G 26}{R27}{2000}

\end{thebibliography}
\end{document}